\documentclass[fleqn,twoside,twocolumn,nofootinbib,showkeys]{revtex4} 

\usepackage[sec,nocpr]{ujp}
\begin{document}
\title[On the phase diagram of the 2d Ising model with frustrating dipole interaction]
{ON THE PHASE DIAGRAM OF THE 2D ISING MODEL WITH FRUSTRATING DIPOLE INTERACTION}%
\author{P.~Sarkanych}
\affiliation{Department for Theoretical Physics, Ivan Franko National University of Lviv}
\address{12 Drahomanov St., 79005 Lviv, Ukraine}
\email{petrosark@gmail.com}
\author{Yu.~Holovatch}
\affiliation{Institute for Condensed Matter Physics of the National Academy of
Sciences of Ukraine}%
\address{79011 Lviv, Ukraine}%
\author{R.~Kenna}%
\affiliation{Applied Mathematics Research Centre, Coventry University}%
\address{Coventry CV1 5FB, UK}%


\autorcol{P.~Sarkanych, Yu.~Holovatch, R.~Kenna}

\setcounter{page}{1}%

\begin{abstract}
Due to intrinsic frustrations of interaction, the 2d Ising model
with competing ferromagnetic short-range nearest-neighbour and
antiferromagnetic long-range dipole interactions possesses a rich
phase diagram. The order of the phase transition from the striped
$h=1$ phase to the tetragonal phase that is observed in this model
has been a subject of recent controversy. We address this question
using the partition function density analysis in the complex
temperature plane. Our results support the second-order phase
transition scenario. To measure the strength of the phase transition
we calculate the values of the specific heat critical exponent
$\alpha$. Along with the space dimension $D$, it appears to depend
on the ratio  of strength of the short-range and long-range
interactions.
\end{abstract}

\keywords{frustrations, phase transition, density of partition
function zeros, critical exponents}

\pacs{64.60.De,75.40.Cx} 

\maketitle

\udk{538.9} 

\section{Introduction} \label{I}

Pattern formation is one of the fascinating phenomena that may be
induced by competing long- and short-range interactions in
many-particle physical systems. To give few examples, a competition
of short-range and long-range dipole interaction leads to a variety
of experimentally observed structures in ultrathin magnetic films on
metal substrates, liquid crystals, polymer films, two-dimensional
electron gases, Langmuir and lipid monolayers, etc. (see
\cite{Giuliani06} for more detailed references). Being a subject of
intensive experimental studies revealing unusual physical effects,
the above systems have important industrial applications. In
particular, the aforementioned ultrathin magnetic films became a subject
of especial interest due to their possible application in creating
high-density storage devices \cite{Pescia90}.

Theoretical insight into peculiarities of pattern formation in the
above systems has been gained by analysing the 2d Ising model with
competing ferromagnetic short-range nearest-neighbour interactions (with strength given by $J$) and
antiferromagnetic long-range dipole interactions (of strength $g$)
\cite{Giuliani06,MacIsaac95,Abanov95,Booth95,Vaterlaus00,Cannas04,Kashuba93,Gleiser02,Rastelli06,Rastelli07,Fonseca12}.
In the framework of this model, the richness of the phase diagram is
attributed to frustrations introduced by the competing character of
ferromagnetic and antiferromagnetic interactions. Analytical approaches,
supported by numerical simulations, show that, depending on
the values of $J$ and $g$, the low temperature phase of this model
is characterized by spin configurations classified as regular and
irregular checkerboards or stripes of different width
$h$\footnote{Here and below $h$ is the stripe width measured in
lattice units.}  with spin oriented in similar direction
\cite{MacIsaac95,Abanov95,Booth95,Cannas04,Vaterlaus00}. Evidence of
modulated phases has been reported for for $J/g$ ratio near the
boundary between striped phases of width $h$ and $h+1$
\cite{Rastelli07}. The above low-temperature magnetic patterns have
much in common with those observed in liquid crystals and the
stripped, modulated, and paramagnetic phases are often referred to
as smectic, nematic and tetragonal ones. In the latter case, the stripe
domains are mutually perpendicular. Moreover, already the mean field
analysis reveals that the domain-wall structure in such films is
similar to 2d liquid crystals \cite{Kashuba93}.

\begin{figure}[tbh]
\begin{center}
\includegraphics[width=\columnwidth]{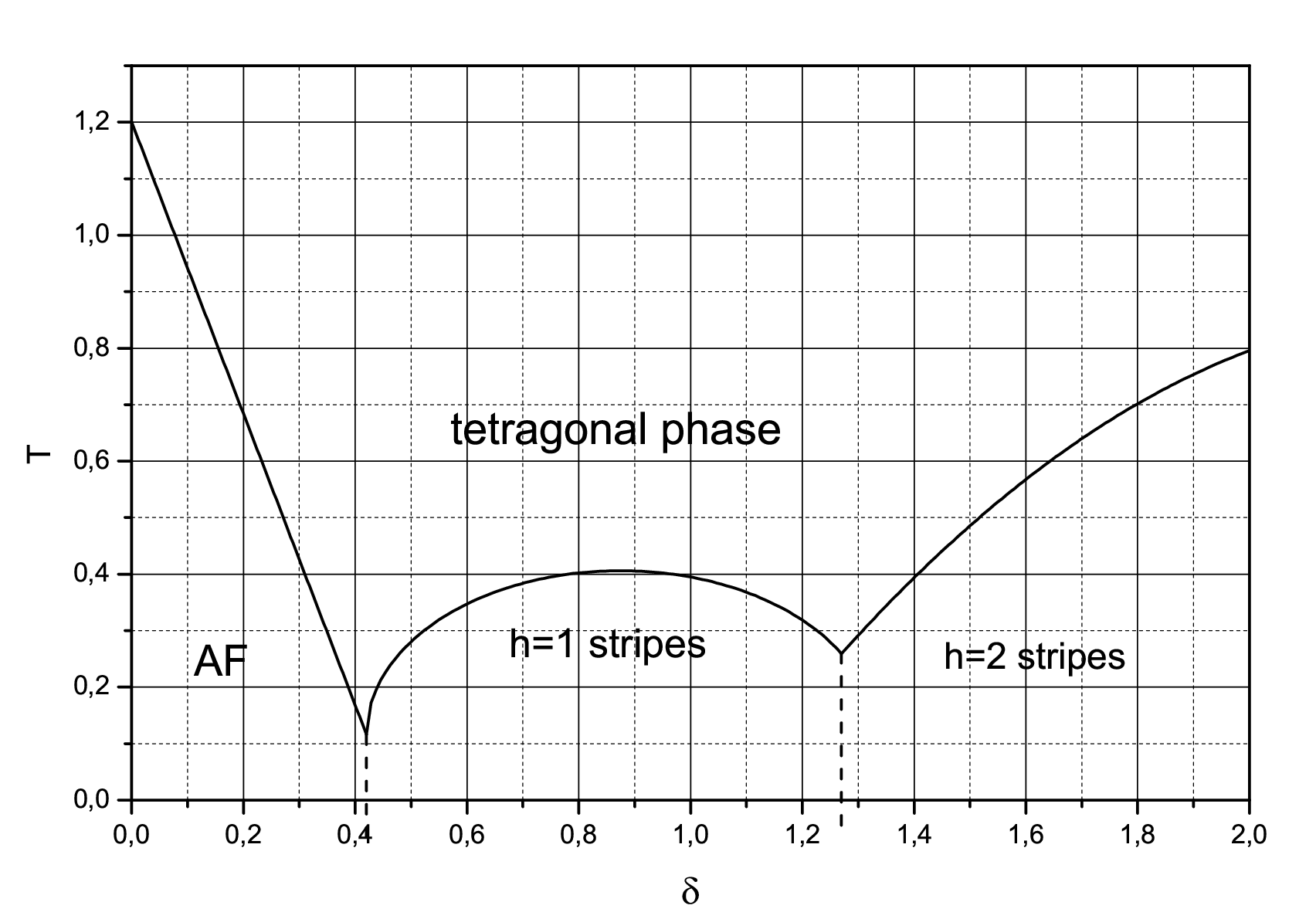}
\caption{\em The phase diagram of the 2d Ising model with competing
nearest neighbour and dipole interactions (sketched from the Refs.
\cite{Fonseca12, Rastelli07}). AF: checkerboard antiferromagnet
phase, $h = 1$, $h = 2$: striped phases. \label{fig1}}
\end{center}
\end{figure}

The  papers cited above agree in general on the classification of
patterns observed in the 2d Ising model with competing
nearest-neighbour and  dipole interactions, however the detailed form of
the phase diagram remains still unclear. In particular, a subject of
recent discussion has been the form of the phase diagram in the
region of low values of temperature $T$ and $\delta=J/g$ (a sketch
of the phase diagram  in this region is shown in Fig. \ref{fig1}).
Whereas the temperature-induced phase transitions from the
antiferromagnetic checkerboard-like phase (AF) and from the striped
$h=2$ phase to the tetragonal phase are of the second and of the
first order, correspondingly, the order of the phase transition
between the striped $h=1$ phase and the tetragonal phase is the
subject of discussion. In particular, MC simulations of Ref.
\cite{Rastelli07} manifest a second order phase transition in
the region $\delta<0.8$ and the first order transition in the region
$0.83 < \delta < 0.88$. As a consequence, a tricritical point has
been predicted in between the two regions. Subsequent MC simulations
of Ref. \cite{Fonseca12} did not observe the conjectured tricritical
point and resulted in the continuous phase transition for all region
of $\delta$ that corresponds to the boundary between the striped
$h=1$ and tetragonal phases (see Fig. 1). The simulations were
supported by the analysis of the partition function zeros in the
complex temperature plane (Fisher zeros) \cite{Fisher68}.

In our study, we will complement analysis of the phase diagram of
the 2d Ising model with competing nearest-neighbour and  dipole
interactions by considering partition function zeros density
\cite{Janke01}. In particular, this will allow to avoid a hyperscaling
assumption while calculating the strength of the phase transition
and directly obtaining the specific heat critical exponent $\alpha$.
The input data for our analysis are provided by the coordinates of
the partition function zeros calculated in \cite{Fonseca12}. The rest
of the paper is organized as follows: in sections \ref{II} and
\ref{III} we describe the model and the method used for its
analysis, our results are presented and discussed in section \ref{IV}.

\section{Ising model with dipole interaction} \label{II}

The Hamiltonian of the 2d Ising model with competing ferromagnetic
nearest-neighbour and antiferromagnetic dipole interactions reads
\begin{equation} \label{1}
H = -\delta \sum_{<i,j>} \sigma_{i} \sigma_{j} + \sum_{i < j}
\frac{\sigma_{i} \sigma_{j}}{r_{ij}^{3}} \, .
\end{equation}
Here, $\delta = J/g > 0$, $J$ and $g$ being strengths of the nearest
neighbour and dipole interactions, correspondingly. The summation is
performed over the sites of the $L\times L$ square lattice. The first
sum in (\ref{1}) spans all pairs of nearest-neighbour Ising spins
$\sigma_i= \pm 1$, while in the second term all pairs of lattice
sites are taken into account. The Ising spins are supposed to be
aligned out of the plane.

In the limiting cases of $J=0$ or $g=0$ (i.e. $\delta$ equals 0 or
$\infty$) the Hamiltonian (\ref{1}) presents a pure dipole interaction
model or the usual Ising model, correspondingly. Both cases are
characterized by a single (antiferro- or ferromagnetic) low
temperature phase and by a continuous second-order phase transition
to the paramagnetic state. Note, that the aforementioned
antiferromagnetic-to-paramagnetic phase transition belongs to the
universality class of the 2d Ising model too
\cite{MacIsaac92,Rastelli06}. The phase behaviour of the model
(\ref{1}) is much more complicated for nonzero $\delta$, as was
briefly described in the introduction. The part of the phase diagram
of the model  in the region of small $\delta$ of $T-\delta$ plane is
sketched in Fig. \ref{fig1}.

\begin{table*}
\renewcommand{\tabcolsep}{0.7cm}
\begin{tabular}{lllllll}
\hline \\
[-0.35cm]
\hline\\
[-0.3cm] $\delta$   &$d\nu$ \cite{Fonseca12}  & $\alpha=2-d\nu$
&$\alpha/\nu$ \cite{Fonseca12}  &
$\alpha=\frac{2\alpha/\nu}{d+\alpha/\nu}$   &$\alpha_{zd}$ \\
\hline
$0.89$       &1.807(70)  &0.193(70)   &0.364(20)       &0.308(17)    &0.194(17)\\
$0.91$       &1.817(68)  &0.183(68)   &0.375(19)       &0.316(16)    &0.191(14)\\
$0.93$       &1.779(61)  &0.221(61)   &0.399(20)       &0.333(17)    &0.221(16)\\
$0.95$       &1.741(53)  &0.259(53)   &0.424(20)       &0.350(17)    &0.255(14)\\
$0.97$       &1.706(46)  &0.294(46)   &0.461(19)       &0.375(16)    &0.292(14)\\
$1.00$       &1.659(37)  &0.341(37)   &0.522(17)       &0.414(14)    &0.349(13)\\
$1.10$       &1.415(25)  &0.585(25)   &0.888(21)       &0.615(15)    &0.5882(84)\\
$1.20$       &1.223(21)  &0.777(21)   &1.496(28)       &0.856(16)    &0.788(17)\\
$1.30$       &1.0093(28) &0.9907(28)  &2.0183(66)      &1.0046(33)   &1.011(13)\\
\hline \\
[-0.35cm] \hline
\end{tabular}
\vskip 4mm \caption{Critical exponents of the 2d Ising model with
competing ferromagnetic nearest-neighbour and antiferromagnetic
dipole interactions for different values of the interaction ratio
$\delta$. Results of Ref. \cite{Fonseca12} obtained by  FSS of
the partition function zeros and of the specific heat, $d\nu$ and
$\alpha/\nu$ are shown in the second and fourth columns,
respectively. The specific heat critical exponents $\alpha$ obtained
via hyperscaling relations from these values are given in the third
and in the fifth columns, correspondingly. The sixth column contains our results
obtained via the partition function zeros density analysis
($\alpha_{\rm zd}$). \label{tab1}}
\end{table*}

Previous analysises of the phase diagram were performed either by numerical
or analytical tools, based on calculation of the partition function
\begin{equation} \label{2}
Z_L (\beta) = {\rm Tr} \exp{(-\beta H)} \, ,
\end{equation}
where $\beta=1/T$ and the trace is performed over all spin
configurations.
The analysis we report upon in this paper
relies on an examination of the partition function behaviour in the
complex $T$ (complex $\beta$) plane. Since the pioneering papers of
Lee and Yang \cite{Lee52} and Fisher \cite{Fisher68} where the
partition function zeros in complex field and complex temperature
planes were studied, this type of analysis became a powerful tool to
study phase transitions in various models. For the model under
consideration (\ref{1}) it has been recently applied in Ref.
\cite{Fonseca12}, where the first zero of the partition function
closest to the origin has been calculated for different values of
interaction ratio $\delta$ at different lattice sizes $L=12 - 72$.
The finite size scaling (FSS) analysis of the zeros' coordinates
allows one to obtain the value of the correlation length critical
exponent $\nu$. The value of $d\nu$ is given in Table \ref{tab1} for
different values of $\delta$. Provided that the hyperscaling
relation $\alpha=2-d\nu$ holds, one can use it to obtain the
specific heat critical exponent $\alpha$. Corresponding
$\alpha(\delta)$ values are quoted in the third column of the Table
\ref{tab1}. Since $\alpha=1$ for  first-order phase transitions, the
obtained values of the exponents $\alpha<1$ serve an evidence for a
second-order phase transition in the considered region of $\delta$.
This result has been further supported by a FSS analysis of the
specific heat, leading to the ratio $\alpha/\nu$ that is quoted in
the fourth column of  Table \ref{tab1} \cite{Fonseca12}. Again,
using the hyperscaling relation one can extract the value of the
$\alpha$ at $d=2$ via: $\alpha=\frac{2\alpha/\nu}{d+\alpha/\nu}$.
The last value is shown as a function of $\delta$ in the fifth
column of the table.

We  note that, although the above-obtained estimates for $\alpha$ support
a continuous phase-transition scenario ($\alpha<1$), they do not agree
numerically. Moreover, the methods used for their determination do
not deliver their direct evaluation, but rather rely on
hyperscaling relations. Therefore, in the forthcoming section we
will use an alternative partition-function-zeros analysis that
allows direct determination of the exponent $\alpha$.

\section{Density of partition function zeros} \label{III}

In our analysis we will use the method of analysing the density of partition
function zeros originally suggested in \cite{Janke01}.
A particular advantage of this method is that it allows to
discriminate between the first- and second-order (as well as higher
order) phase transitions as well as to measure the strength of first-
and second-order phase transitions in the form of the latent heat and
critical exponents. Below we briefly describe the main steps of the
partition function density analysis. Provided that the zeros of the
partition function of the model (given by (\ref{2}) in our
particular case) in the complex plane are known, one can write it in
the factorized form
\begin{equation} \label{3}
 Z_L(z) = A(z) \prod_{j}{\left(z-z_j(L)\right)}
\quad ,
\end{equation}
where $z$ stands generically for an appropriate function of complex
temperature (in the Fisher case) or complex field (in the Lee-Yang
case), $L$ is the linear extent of the lattice and $A(z)$ is a
smooth function that never vanishes. The free energy density follows
as
\begin{eqnarray} \label{4}
 f_L(z) &=& \frac{1}{L^d}\ln{Z_L(z)} \\ \nonumber
 &=&
\frac{1}{L^d}\left(\ln{A(z)}+\sum_{j}{\ln{(z-z_j(L))}}\right)
 \, .
\end{eqnarray}
The first term on the right contributes only to the regular part of
thermodynamic functions and will be dropped henceforth. The
remainder, which will be referred to as $f_L^{\rm{s}}(z)$, gives
rise to singular behaviour.

It is suitable to parameterize the zeros by
\begin{equation} \label{5}
z = z_c + r \exp{(i \varphi)} \, ,
\end{equation}
where $z_c$ is a critical point coordinate. Let us define the
density of zeros as
\begin{equation}
 g_L(r) = L^{-d} \sum_{j} \delta(r - r_j(L)) \, ,
 \label{J}
\end{equation}
with $z_j=z_c+r_j \exp{(i \varphi)}$.  Subsequently, the free energy
and the cumulative distribution function of zeros are defined as
\begin{equation}
 f_L^{\rm{s}}(z) = \int_{0}^{R}{g_L(r)\ln{(z-z_c - re^{i \varphi})} dr}
 + {\rm{c.c.}}
\quad ,
\label{R}
\end{equation}
\begin{equation}
 G_L(r)=\int_0^r{ g_L(s) d s}=
 \begin{cases}
 =\cfrac{j}{L^d} \quad \quad {\rm{if}} \quad r \in (r_j,r_{j+1})\\
 =\cfrac{2j-1}{2L^d} \quad \quad {\rm{if}}\quad r=r_j
\end{cases}
\label{8}
\end{equation}
where c.c. means complex conjugate and $R$ is some appropriate
cutoff. In the thermodynamic limit and for  first order phase
transitions, Lee and Yang already have shown \cite{Lee52} that the
density of zeros has to be non--zero crossing the real axis. This
corresponds to the cumulative distribution of zeros
\begin{equation}
 G_\infty(r) = g_\infty(0) r
+ br^{w+1} + \dots
\quad ,
\label{1st}
\end{equation}
where the slope at the origin is related to the latent heat
(or magnetization) via
\begin{equation}
  \Delta e \propto g_\infty(0)
\quad .
\label{1stl}
\end{equation}
Furthermore, it has been shown (see \cite{Abe67, Suzuki67}) that the
necessary and sufficient condition for the specific heat at
second-order phase transitions to have the leading critical behaviour
$C \sim t^{-\alpha}$, is
\begin{equation}
 G_\infty(r) \propto  r^{2-\alpha}
\quad .
\label{2nd}
\end{equation}

The above survey leads to the conclusions that a plot of $G_L(r_j)
= (2j-1)/2L^d$ against $r_j(L)$ should: (i) go through the origin,
(ii) display $L$-- and $j$-- collapse, and (iii) reveal the order
and strength of the phase transition by its slope near the origin.
In the next section we will give results of an analysis of
corresponding plots for the partition function (\ref{2}) of the
model with the Hamiltonian (\ref{1}).

\section{Results and conclusions} \label{IV}

\begin{figure}[th]
\begin{center}
\includegraphics[width=\columnwidth]{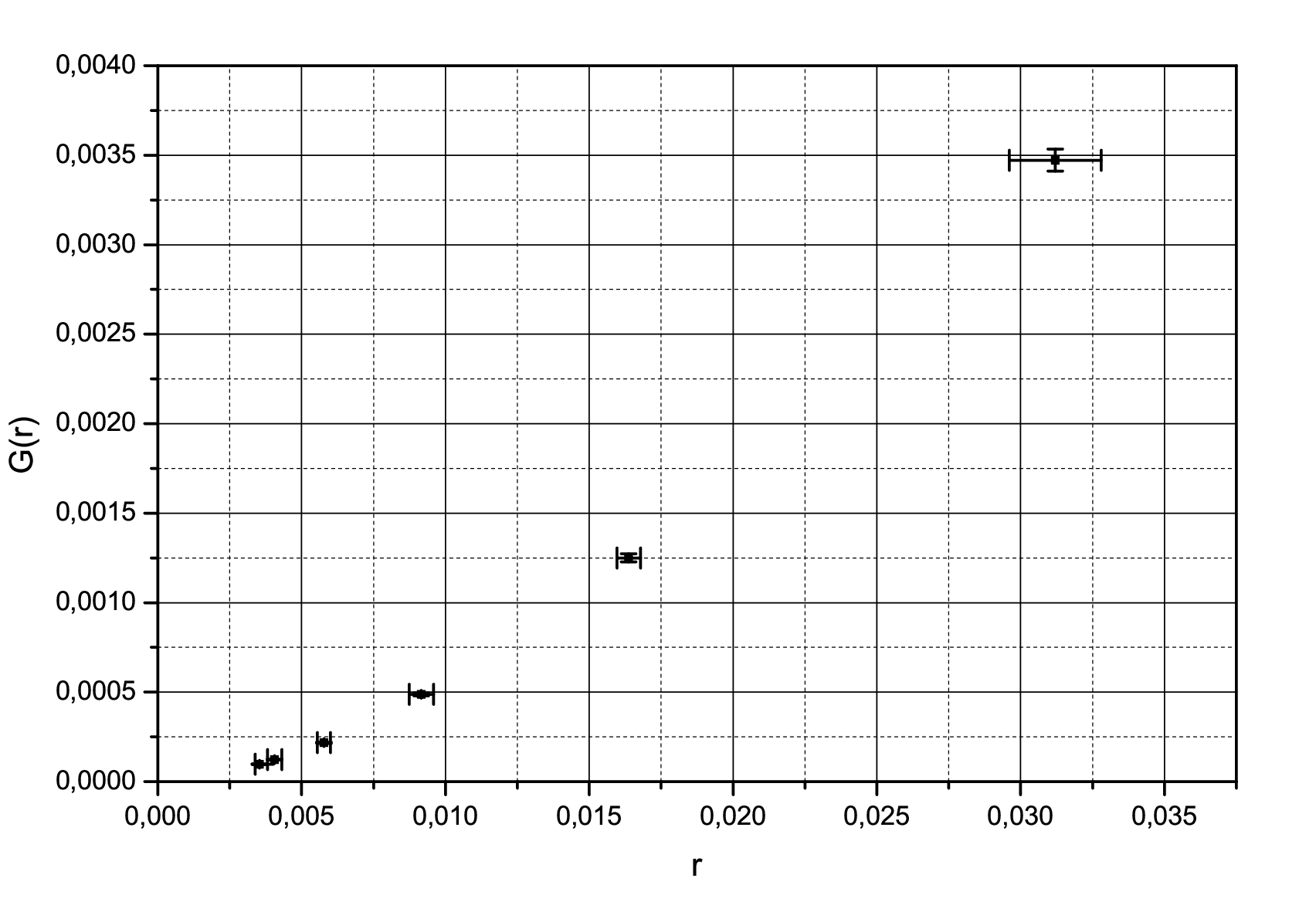}
\caption{{Cumulative density function $G(r)$ for $\delta = 1$.
\label{fig2}}}
\end{center}
\end{figure}

As it has been stated in the Introduction, the input data for our analysis are provided by the coordinates of
the partition function zeros calculated in \cite{Fonseca12} for
different $\delta$ and $L$.
From these, we calculate $G(r)$
dependence of the cumulative density function (\ref{8}) for different
values of $\delta$. Obtained in this way, a  typical $G(r)$ plot is
shown in Fig. \ref{fig2} for $\delta=1$. Subsequently, the set of
functions $G(r)$, for every value of $\delta$, is fitted  with
$G(r)=a r^{2-\alpha}+b$. The resulting values of the specific heat
critical exponent are listed in the last column of Table \ref{tab1},
$\alpha_{\rm zd}$.

There are several conclusions one can make comparing the data for
the specific heat exponents from Table \ref{tab1}. First of all it
is worth noting that the results obtained by three different
techniques: (i) FSS of partition function zeros (third column of the
table, obtained by hyperscaling relation from data of Ref.
\cite{Fonseca12}), (ii) FSS of the specific heat \cite{Fonseca12}
(fifth column of the table) and (iii) density of partition function
zeros analysis (last column of the table, our data) give value of
$\alpha<1$ up to $\delta<1.3$. Recalling that that $\alpha=1$ serves
as evidence for a first-order phase transition (cf. Eqs. (7) and
(9)) one can conclude that the transition from the striped $h=1$ to
the tetragonal phase (see the phase diagram Fig. \ref{fig1}) occurs
through a continuous transition scenario. For the values of $\delta
\geq 1.3$  all three approaches predict a first-order phase
transition, leading to the conclusion that the tricritical point is
located in region $1.2<\delta<1.3$.

\begin{figure}[th]
\begin{center}
\includegraphics[width=\columnwidth]{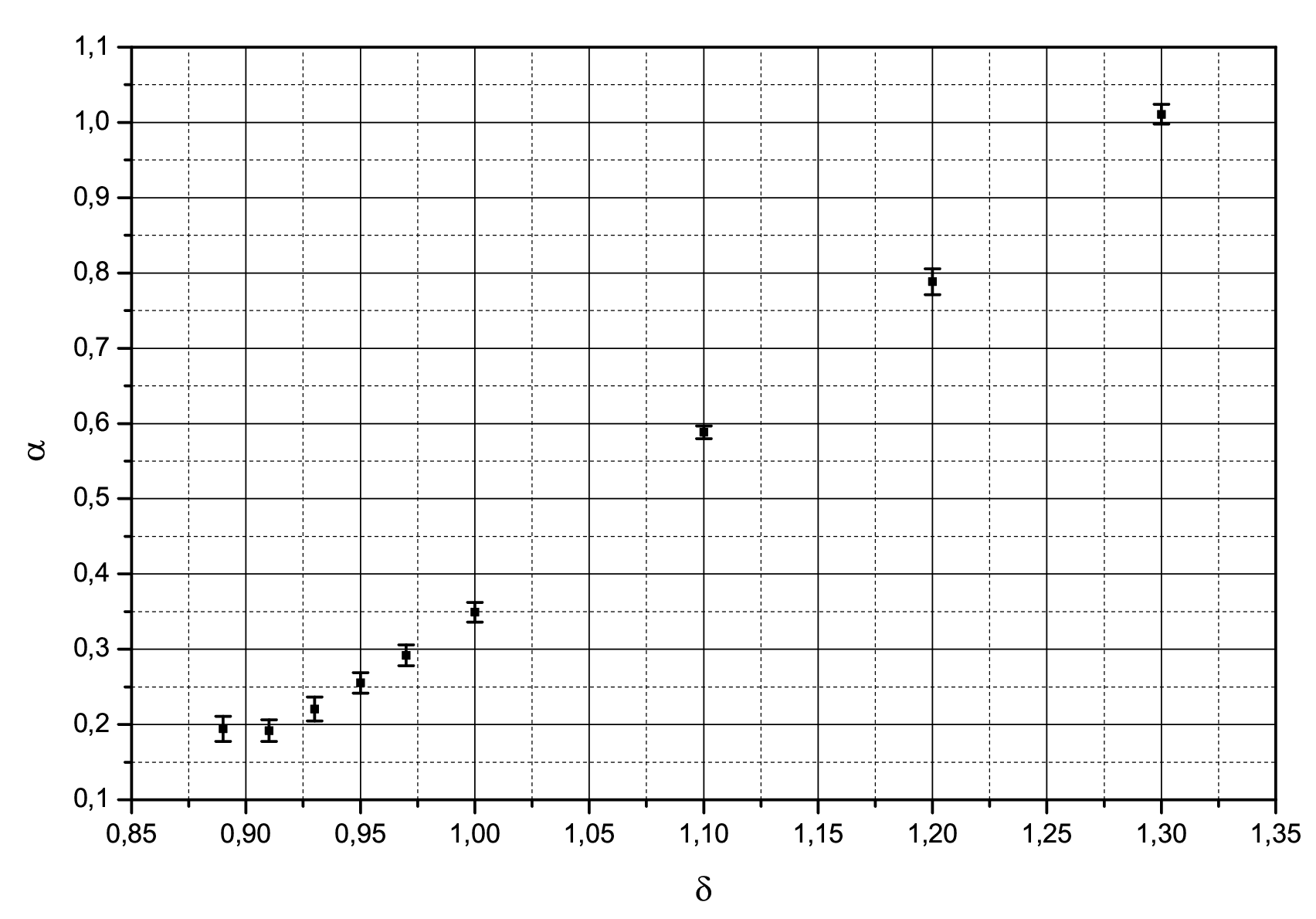}
\caption{Dependency of the critical exponent $\alpha$ calculated in
this study ($\alpha_{\rm zd}$) on the interaction parameter
$\delta$. \label{fig3}}
\end{center}
\end{figure}

All three approaches deliver $\delta$-dependent values of the
critical exponents, making $\delta$ along with the space dimension
$d$ the global variable that defines the universality class.
Dependency of the critical exponent $\alpha$ calculated in this
study ($\alpha_{\rm zd}$) on the interaction parameter $\delta$ is
shown in Fig. \ref{fig3}. Let us note however, that the numerical
values of the exponents obtained via different approaches differ. In
particular, results of the FSS analysis of partition function zeros
(third column of the table) are in good agreement with the
analysis of the density of partition function zeros (sixth column of the
table). But they  essentially differ from the results obtained on the basis of a
FSS analysis of the specific heat behaviour \cite{Fonseca12}. This
result calls for further investigation.

This work is supported in part by FP7 Marie Curie Action grants
PIRSES-GA-2011- 295302-SPIDER and PIRSES-GA-2010-269139 -
DCP-PhysBio. P.S. acknowledges useful discussion during the young
scientists conference at the Bogolyubov Institute for Theoretical
Physics (Kyiv, Ukraine, December 24-27, 2013) where this work was
presented.

\end{document}